\documentclass[%
 reprint,
 amsmath,amssymb,
 aps,
]{revtex4-1}

\usepackage{array}
\usepackage{epsfig}
\usepackage{graphicx}
\usepackage{xy}
\usepackage{siunitx}
\usepackage{dcolumn}
\usepackage{bm}
\usepackage{xcolor}
\usepackage{soul}
\setstcolor{red}
\usepackage{ulem}
\newcommand\redout{\bgroup\markoverwith
{\textcolor{red}{\rule[.5ex]{2pt}{0.4pt}}}\ULon}
\usepackage{makecell}
\usepackage{comment}
\usepackage[version=4]{mhchem}
\newcommand{\psf}{PS$_4$}
\newcommand{\Argy}{Li$_6$PS$_5$Cl}

\newcommand{\symcell}{Li$|$Li$_6$PS$_5$Cl$|$Li}
\newcommand{\product}{Li$_2$S, Li$_3$P, and LiCl}

\usepackage{adjustbox}
\usepackage{multirow}
\usepackage{changepage}
\newcolumntype{L}[1]{>{\raggedright\let\newline\\\arraybackslash}m{#1}}
\newcolumntype{C}[1]{>{\centering\let\newline\\\arraybackslash}m{#1}}
\newcolumntype{R}[1]{>{\raggedleft\let\newline\\\arraybackslash}m{#1}}

\usepackage{xr-hyper}
\usepackage{hyperref}

\usepackage{xr}
\begin{document}

%\title{Reaction and ion transport at electrode-electrolyte interface in lithium metal solid-state batteries}

\title{Coupled reaction and diffusion governing interface evolution in solid-state batteries}

\author{Jingxuan Ding,$^{1 \ast}$ Laura Zichi,$^{1}$ Matteo Carli,$^{1}$ Menghang Wang,$^{1}$ Albert Musaelian,$^{1}$ Yu Xie,$^{1}$ Boris Kozinsky$^{1,2 \ast}$}

%\begin{affiliations}
\affiliation{
\normalsize{$^{1}$Harvard John A. Paulson School of Engineering and Applied Sciences, Harvard University, }\normalsize{Cambridge, MA, USA}\\
\normalsize{$^{2}$Robert Bosch LLC Research and Technology Center, }\normalsize{Watertown, MA, USA}\\
\normalsize{$^\ast$To whom correspondence should be addressed; E-mail: jingxuanding@seas.harvard.edu, bkoz@seas.harvard.edu}
}
%\end{affiliations}

\begin{abstract}
\noindent 
Understanding and controlling the atomistic-level reactions governing the formation of the solid-electrolyte interphase (SEI) is crucial for the viability of next-generation solid state batteries. However, challenges persist due to difficulties in experimentally characterizing buried interfaces and limits in simulation speed and accuracy. We conduct large-scale explicit reactive simulations with quantum accuracy for a symmetric battery cell, {\symcell}, enabled by active learning and deep equivariant neural network interatomic potentials. To automatically characterize the coupled reactions and interdiffusion at the interface, we formulate and use unsupervised classification techniques based on clustering in the space of local atomic environments. Our analysis  reveals the formation of a previously unreported crystalline disordered phase, Li$_2$S$_{0.72}$P$_{0.14}$Cl$_{0.14}$, in the SEI, that evaded previous predictions based purely on thermodynamics, underscoring the importance of explicit modeling of full reaction and transport kinetics. Our simulations agree with and explain experimental observations of the SEI formations and elucidate the Li creep mechanisms, critical to dendrite initiation, characterized by significant Li motion along the interface. Our approach is to crease a digital twin from first principles, without adjustable parameters fitted to experiment. As such, it offers capabilities to gain insights into atomistic dynamics governing complex heterogeneous processes in solid-state synthesis and electrochemistry.

\end{abstract}

\pacs{Valid PACS appear here}

\maketitle

%%%%%%%%%%%%%%%%%%%%% sections not needed for Nature Journals 
\section*{Introduction}
%%%%%%%%%%%%%%%%%%%%% sections not needed for Nature Journals 
\noindent All-solid-state batteries (ASSBs) are attracting significant attention due to their enhanced safety and higher theoretical capacity with lithium metal anodes compared to traditional lithium-ion batteries \cite{janek2016solid,liu2019pathways}. The key to high performance in ASSBs are solid electrolytes with high lithium ion conductivity \cite{kamaya2011lithium,kato2016high,adeli2019boosting}, electrochemically stable interfaces with the electrode at its operating potential \cite{yu2016synthesis,zheng2018review}, and mechanical stability \cite{famprikis2019fundamentals,banerjee2020interfaces,zhang2023homogeneous}. The solid electrolyte interphase (SEI), formed via the reaction  of the  solid electrolyte  with lithium metal, critically affects battery performance due to the increased ionic and electronic resistance of reaction products \cite{janek2016solid,famprikis2019fundamentals,schwietert2020clarifying}. Lithium dendrite growth can lead to internal shorting of the cell \cite{kasemchainan2019critical,ning2023dendrite}, while the formation of inactive lithium during cycling results in the loss of capacity and Coulombic efficiency \cite{fang2019key,chen2021new,jin2021rejuvenating}, making the suppression of both phenomena essential for long-cycle life. 

Electrochemical interface stability has been previously investigated using first-principles thermodynamic calculations \cite{zhu2015origin,richards2016interface}. However, there are notable discrepancies with the experimental measurements \cite{kamaya2011lithium,kato2016high}, where computational predictions underestimate stability windows of lithium-electrolyte interfaces \cite{tan2019elucidating,chen2019approaching,schwietert2020clarifying,xiao2020understanding}. The limitations of previous approaches arguably arise due to the lack of consideration for kinetic phenomena, such as interdiffusion, nucleation and growth of metastable phases, which are known to be as important in controlling solid state reactions as thermodynamic driving forces \cite{schmalzried1974solid}.
To suppress  the decomposition, recent experimental works propose multilayer architectures, where a more-stable electrolyte is inserted between he elecrode and a less-stable higher-mobility electrolyte \cite{ye2021dynamic,ye2024fast}. These architectures stabilize the heterogeneous electrolyte-electrode interfaces, prevent lithium dendrite penetration, and enhance cycling performance in ASSBs. However, such strategies increase the complexity and cost of processing, and ideally passivation product layers can form in-situ. Gaining the ability to control and predict both thermodynamics and kinetics of SEI formation and reaction passivation demands a deeper exploration of the underlying mechanisms.

Atomistic-level understanding of the electrochemical reactions forming the SEI remains challenging due to the difficulty of in-situ characterization of buried interfaces. Techniques like X-ray photoelectron spectroscopy (XPS), scanning/transmission electron microscopy, Raman spectroscopy, and time-of-flight secondary-ion mass spectrometry can provide partial insights into the SEI composition, and morphology evolution \cite{pang2017vivo,wenzel2018interfacial,tan2019elucidating,narayanan2022effect,otto2022situ,alt2024quantifying}, but do not reveal atomistic level mechanistic details. Identification of amorphous phases and nanoscale multi-component structure of interfacial product layers at sufficient spatial and temporal resolution is a persistent challenge in the field. More generally, similar challenges preclude progress in understanding solid-state synthesis mechanisms and processes in surface degradation in lubricants, catalysts, and epitaxial growth of materials.

\begin{figure*}[ht!]
\includegraphics[width=1\linewidth,trim={0cm 0cm 0cm 0cm},clip]{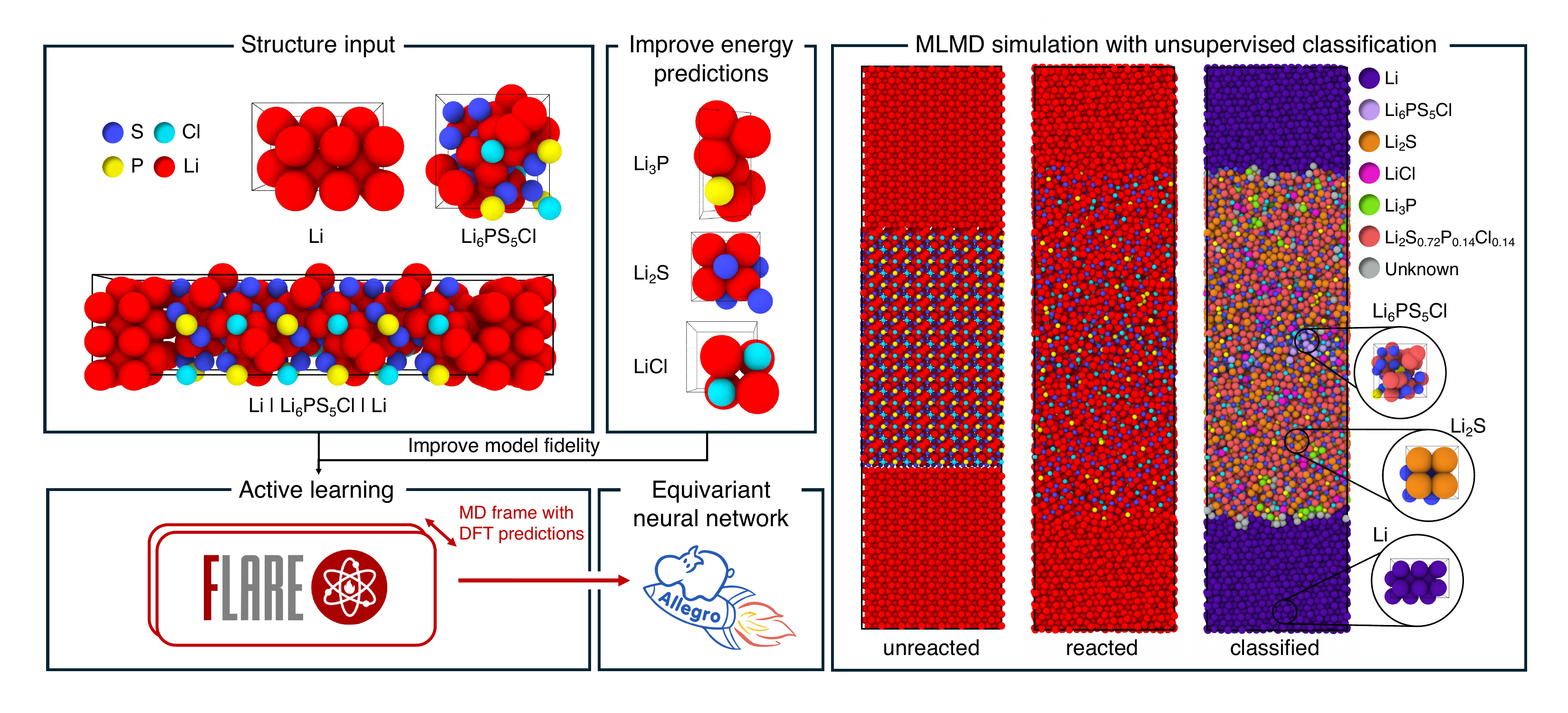}
\caption{\textbf{Schematic workflow of data generation, model construction, and MD simulation}: (1) collecting training data via on-the-fly active learning, (2) constructing an equivariant neural network model, (3) iterating on the machine learning interatomic potential fidelity and incorporating additional data for accurate energy predictions for reaction products, (4) conducting final production runs using large-scale machine learning molecular dynamics, and (5) unsupervised structure identification to automatically characterize the rapid reaction progression.
} 
\label{fig:workflow}
\end{figure*}

Explicit atomistic computational modeling can significantly elucidate the interfacial processes but faces a trade-off between accuracy and cost. Quantum mechanical methods based on density functional theory (DFT), such as \textit{ab initio} molecular dynamics (AIMD) have been applied to reactive systems. However, AIMD simulations are limited to a few hundred atoms and a few hundred picoseconds. This restricts such investigations to only modeling the initial degradation on a small scale, which does not extend to the actual product formation \cite{cheng2017quantum,golov2021molecular,luo2022nanostructure,golov2023unveiling,hao2024first}. 
While classical MD offers efficiency through parameterized interatomic potentials for reactions \cite{van2001reaxff,senftle2016reaxff}, sufficiently accurate interatomic potentials for complex heterogeneous processes has until recently remained challenging. 
Recent advances in machine learning interatomic potentials (MLIPs) allow for quantum accuracy in large-scale simulations to investigate reaction kinetics \cite{behler2007generalized,bartok2010gaussian,shapeev2016moment,wang2018deepmd,xie2018crystal,chen2019graph,batzner2023,musaelian2023learning,vandermause2020fly,vandermause2022active,xie2023uncertainty}. MLMD studies have explored diffusion in mixed amorphous LiF/Li$_2$CO$_3$ \cite{hu2022impact} and the resistive Li-ion transport in selected decomposition products from Li$_6$PS$_5$Cl and Na$_3$PS$_4$ \cite{wang2022resistive}. Larger-scale simulations of symmetric cells have focused on initial reaction stages, e.g., the decomposition of Na$_3$PS$_4$ \cite{bekaert2023assessing}, the SEI formation in Li$|$Li$_3$PS$_4$$|$Li \cite{ren2024visualizing} and two-step growth mechanism in {\symcell} \cite{chaney2024two}. However, these investigations did not examine the long-time coupled mechanical, kinetic, and thermodynamic phenomena that govern the electrode-electrolyte reactions. In particular, they overlooked the critical need to adapt the ML model to account for the formation of unanticipated products and to thus ensure chemical transferability and accuracy through the entire process. Our previous work highlights the importance of careful incorporation of chemical potential into training MLIP models across wide composition space \cite{goodwin2024transferability}. A full description of the SEI evolution must rely on a model that can adapt to the formation of kinetically stabilized non-crystalline and non-stoichiometric reaction products. %A full description of the SEI evolution must rely on a model that can adapt to phase transformations in Li metal as well as the formation of kinetically stabilized non-crystalline and non-stoichiometric reaction products. 

In this work, we employ on-the-fly Bayesian active learning based on Gaussian Process regression (FLARE) \cite{vandermause2020fly,vandermause2022active,xie2023uncertainty} and deep equivariant neural network interatomic potentials (Allegro) \cite{musaelian2023learning} to develop a MLIP and perform large-scale, long-time explicit reactive MLMD simulations of symmetric battery cells {\symcell} with up to 3 million atoms. Our simulations reveal rapid reaction and interdiffusion at interfaces, formation of polycrystalline SEI of a few nanometers in thickness, and Li creep relevant for dendrite formation. By deploying an adaptive unsupervised density-peak clustering of high-dimensional local environment descriptors, we identified the formation of a previously unknown crystalline disordered phase, Li$_2$S$_{0.72}$P$_{0.14}$Cl$_{0.14}$, in the SEI. We predict that this phase inhibits electron transport while maintaining limited Li$^+$ ion transport via defect migration and grain boundaries, which explains experimental observations. Finally, million-atom simulations reveal distinct ionic transport characteristics between Li$_2$S$_{0.72}$P$_{0.14}$Cl$_{0.14}$ and Li$_2$S, where thin Li$_2$S coating layers completely prevent the Li-ion transport and the reactions between Li and {\Argy}.
More generally, this study offers a promising approach for accelerating the simulations and automating the analysis of atomistic mechanisms in complex heterogeneous processes, such as solid-state synthesis reactions and electrochemical degradation. 

\begin{figure*}[ht!]
\includegraphics[width=0.95\linewidth,trim={0cm 0cm 0cm 0cm},clip]{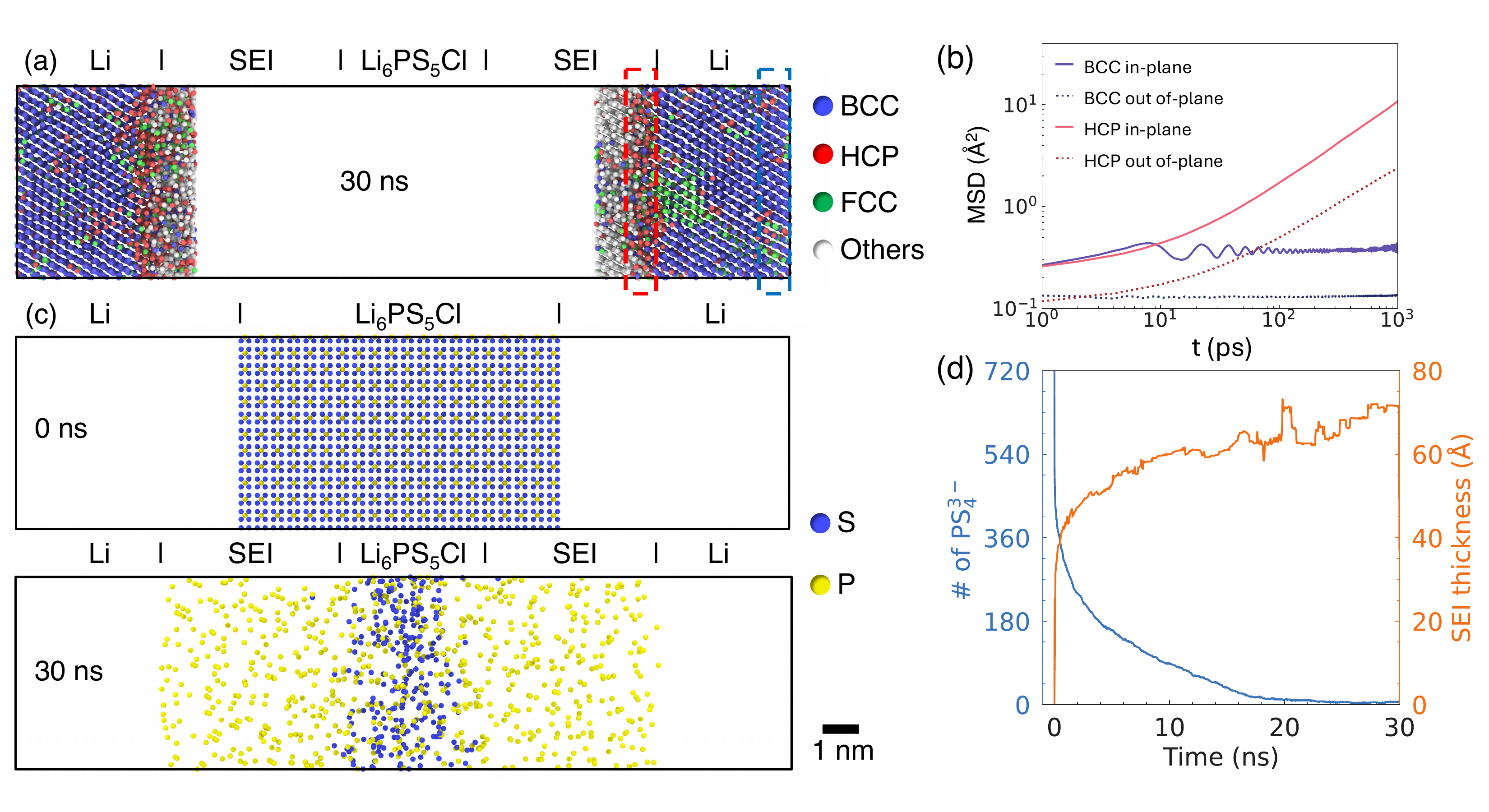}
\caption{\textbf{Li metal evolution and {\Argy} reduction}.
(a) Snapshot of a 21,120-atom cell at 30\,ns, color coded using polyhedral template matching \cite{larsen2016robust}, with blue, red, green, and gray representing body-center cubic (BCC), hexagonal close packed (HCP), face-center cubic (FCC), and other crystal structures. 
(b) Window-averaged mean square displacements (MSD) for BCC and HCP Li within slabs [dashed boxes in (a)]. Li atoms belonging to the BCC and HCP phases are identified based on MD frames at 5, 10, 15, 20, and 25\,ns. The MSD is calculated over the subsequent 2\,ns, averaging over all atoms. 
(c) Snapshots of MLMD frames at 0 and 30\,ns highlighting the PS$_4$ tetrahedra and phosphorus atoms, illustrating a layer-by-layer decomposition of the {\Argy}. Li, Cl, and S not in PS$_4$ tetrahedra are removed for clarity. 
(d) The number of intact PS$_4$ tetrahedra (blue) as a function of time, illustrating the amount of unreacted electrolyte. SEI thickness is shown in orange.
} 
\label{fig:Reaction_mechanism}
\end{figure*}

\section*{Results}
\subsection*{Training an accurate model for interfacial reaction}
An accurate model for simulating chemical reactions of lithium metal with the solid electrolyte {\Argy} has been developed by integrating efficient data collection through on-the-fly active learning and equivariant neural network interatomic potential with state-of-the-art accuracy, as depicted in the workflow in Fig.\,\ref{fig:workflow}. The detailed training process, MLIP construction, and model validation are provided in the Methods section and the Supplementary Information. We collected 7,439 frames from the active learning process, supplemented with 400 frames from reactive MD simulations, 25\,ns to stabilize SEI in a 228-atom cell using the initial potential, 
and an additional 300 frames from AIMD trajectories of thermodynamically predicted reaction products -- {\product} -- yielding a total of 8,139 frames. We find that the model robustness was significantly improved after including training data for the expected reaction products. Initially, without these data, the energy predicted by the model for the expected products has mean absolute error (MAE) and root mean square error (RMSE) as high as 331\,meV/atom comparing to DFT, e.g., in Li$_3$P (Fig.~S4), which could lead to the formation of spurious phases. Upon augmenting the Allegro model with these additional data, the MAE and RMSE were reduced to below 3\,meV/atom (Fig.~S3). While MAE and RMSE are good metrics for benchmarking MLIP models, they do not fully capture the model's ability to accurately predict material properties and reactions. 
Previous simulations of battery interface reactions \cite{ren2024visualizing,chaney2024two} relied on models trained only on reactants, and as a result displayed non-physical behavior including both the instability of crystalline Li metal at 300\,K \cite{chaney2024two} and existence of crystalline Li at 600\,K above its melting point \cite{ren2024visualizing}. 
Our refined reactive model achieves excellent accuracy across an extensive set of benchmarks, including mechanical properties of bulk Li, bulk electrolyte diffusion, accurate energy vs volume changes for reactants, and consistent predictions of energy, forces, and stress, as shown in Table.~S2, Fig.~S1-~S3. It also correctly captured the melting of lithium above its melting point (Fig.~S5). This careful validation is necessary to achieve a reliable model for describing heterogeneous reactions and the resulting competition between metastable phases and structural polymorphs. 

Using the MLIP model constructed as detailed above, we performed MLMD simulations on multiple configurations of the {\symcell}. A 21,120-atom cell was simulated for 35\,ns to investigate the SEI formation and examine the conventional thermodynamically expected pathway Li$_6$PS$_5$Cl + 8Li $\rightleftharpoons$ 5Li$_2$S + Li$_3$P + LiCl \cite{zhu2015origin,wenzel2018interfacial}. A 120,960-atom cell was simulated for 10\,ns to capture the effects of length scale and on chemo-mechanics of the SEI formation (0.07\,$\mu$m long). Two three-million-atom cells (0.26\,$\mu$m long), with and without Li$_2$S layers, were simulated up to 1\,ns to examine the role of crystalline Li$_2$S in the SEI and to establish the scalability of our computational approach, demonstrating the largest battery interface simulation to date. Notably, the simulation speed exhibited nearly linear weak scaling with the number of GPUs as the number of atoms increased proportionally, as detailed in Ref.~\cite{musaelian2023learning}.

\subsection*{Reaction mechanism between Li and {\Argy}: \\the origin of lithium creep}
ASSBs with lithium metal anodes face significant challenges in their practical application, primarily due to issues such as dendrite growth and inactive Li formation in SEI during cycling, where the latter was attributed as the primary reason for performance decay \cite{fang2019key,chen2021new,jin2021rejuvenating}.
Understanding the Li metal evolution in the initial ASSBs assembly stage is essential for the design of a more stable interface. One of the goals of our simulations is to study the  evolution of the Li structure, and we uncover the microscopic origins of Li creep as a diffusive in-plane motion arising from the buffer Li atoms participating in the interface reaction. The deformation of Li due to chemical reactions was suggested experimentally by the observation of reduced stack pressure under open circuit conditions \cite{lee2021stack}. It was suggested that pressure-driven creep deformation of Li metal compensated for voids at the Li/electrolyte interface during cycling \cite{wang2019characterizing,kasemchainan2019critical}, pointing to non-trivial coupling between electrochemical reaction and phase evolution of bulk Li. Previous simulations of Li insertion into a fixed Li$_2$O surface demonstrated the structural transition near the interface from disordered to mixed HCP, FCC, and eventually BCC phases of Li, where higher-energy intermediate phases form before reaching thermodynamic stability \cite{yang2023lithium}. However, the use of a fixed surface neglects the mechanical response and dynamic reconstruction of the interface, which are critical in real systems.

Upon contact between Li metal and the electrolyte, our simulations at 300\,K revealed that a prominent fast reaction occurred already within a few picoseconds, forming an amorphous product region at the boundary. While most of the Li metal remained in the BCC phase, the FCC phase appeared transiently, which is macroscopically expected at pressures above 6.9\,GPa at 300\,K \cite{olinger1983lithium}. This indicates large local internal stress in the reaction zone, resulting from the rapid evolution of the chemical potential as the reaction begins. One example of this behavior is the formation of a transient FCC stripe before 4\,ns, which subsequently reverted to the BCC phase (Fig.~S7). Interestingly, this was accompanied by transient appearance of nanoscale crystalline domains of composition close to Li$_2$S, that appeared at 1\,ns and persisted [Fig.~S8 (b)].

Additionally, one or two layers of BCC Li transformed into a mixture of BCC, FCC, and HCP phase, potentially accommodating transient reaction-induced stresses and serving as a buffer layer between Li metal and the SEI for subsequent diffusion or SEI formation. This buffer layer persisted throughout the simulation, as shown in Fig.~\ref{fig:Reaction_mechanism} and ~S7. By calculating the window-averaged mean square displacements (MSD), we observed different behaviors in lithium metal anode and this buffer layer. Using HCP Li as an example. BCC Li showed no diffusion, only vibrations with small amplitudes. In contrast, HCP Li near the boundary exhibited considerably larger MSD, corresponding to in-plane diffusion along the interface, as shown in Fig.~\ref{fig:Reaction_mechanism} (b). It is encouraging that our large-scale MLMD simulations directly capture these subtle and difficult to observe interfacial Li creep phenomena without any prior information. The out-of-plane MSD for HCP Li also show diffusive motions as these HCP Li further diffuse into the SEI.

On the electrolyte side of the reaction, we saw that the crystalline {\Argy} was steadily consumed in a layer-by-layer manner. We visualized the formation of the {\Argy}$|$SEI boundary by analyzing the distribution of {\psf} tetrahedra, as shown in Fig.~\ref{fig:Reaction_mechanism} (c) and S9. We specifically counted the number of ``intact" {\psf} tetrahedra, defined as those in which all four nearest neighbors of P are S atoms within a cutoff distance of 2.6\,{\AA}, to represent the unreacted {\Argy}, as illustrated in Fig.~\ref{fig:Reaction_mechanism} (d). A sharp decrease in the number of intact tetrahedra was observed within the first nanosecond, accompanied by the decrease in the system volume and total energy (see Fig.~S6). Due to the relatively stable temperature (300$\pm$10\,K) maintained by the thermostat, this energy drop was primarily attributed to a decreased potential energy as the reaction formed thermodynamically favorable products. The corresponding thickness of the SEI was defined by the distance between the boundary of intact {\psf} tetrahedra and the furthest z-coordinates of S, P, and Cl that diffuse into the SEI.  After 30 ns of our MD simulation, we observed the formation of an SEI, comprising both amorphous and crystalline regions, that reached a thickness of approximately 7\,nm on each side of the symmetric cell.  
The formation pattern was consistent across all three configurations simulated, which varied in the initial fraction of Li metal relative to the amount of {\Argy}, and showed similar resulting SEI thickness (see Fig.~S17). The SEI growth slows down substantially after 30\,ns, but does not stop, subject to transport limitations by the slowest-diffusing reacting components, namely P, S, and Cl atoms.

This estimated SEI layer thickness matches the direct observation of SEI between a single Li dendrite and {\Argy} using high-resolution transmission electron microscopy \cite{luo2022nanostructure}. However, this value is lower than the estimation obtained from time-of-flight secondary-ion mass spectrometry or electrochemical titration, which reported thicknesses of a few hundred nanometers after one week \cite{otto2022situ,aktekin2023sei,alt2024quantifying}. By directly synthesizing the SEI via a reaction between micro-sized lithium and {\Argy}, Alt \textit{et al.} estimated a thickness of 126\,nm after one day from conductivity measurements \cite{alt2024quantifying}.

\begin{figure}[t]
\includegraphics[width=1\linewidth,trim={0cm 0cm 0cm 0cm},clip]{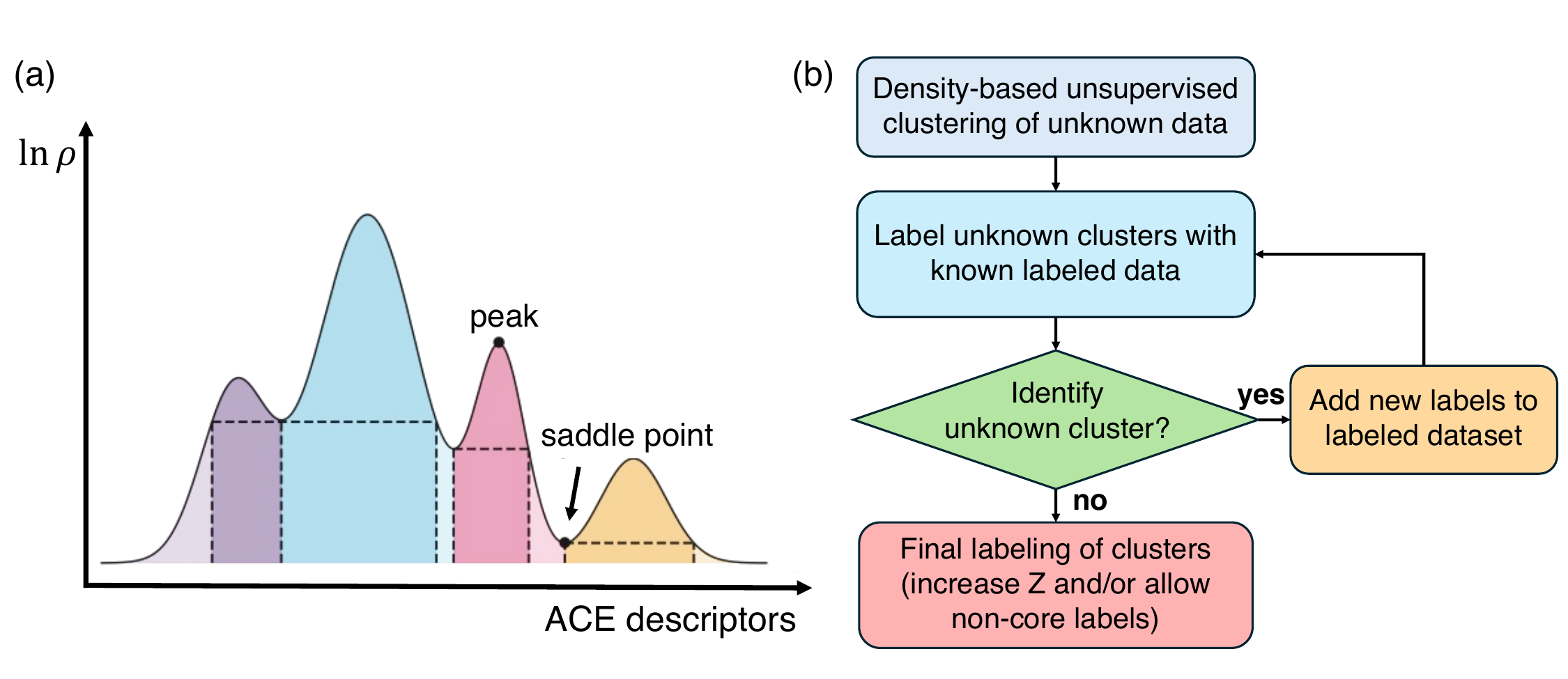}
\caption{\textbf{Unsupervised characterization of interfacial reactions.}(a) 1D schematic of density estimation of atomic cluster expansion (ACE) descriptors. Dark shaded colors represent core regions, and light shaded colors represent non-core regions. (b) Workflow of the unsupervised classifier highlighting how a new phase is discovered.}
\label{fig:DPA_workflow} 
\end{figure}

\begin{figure*}
\includegraphics[width=1\linewidth,trim={0cm 0cm 0cm 0cm},clip]{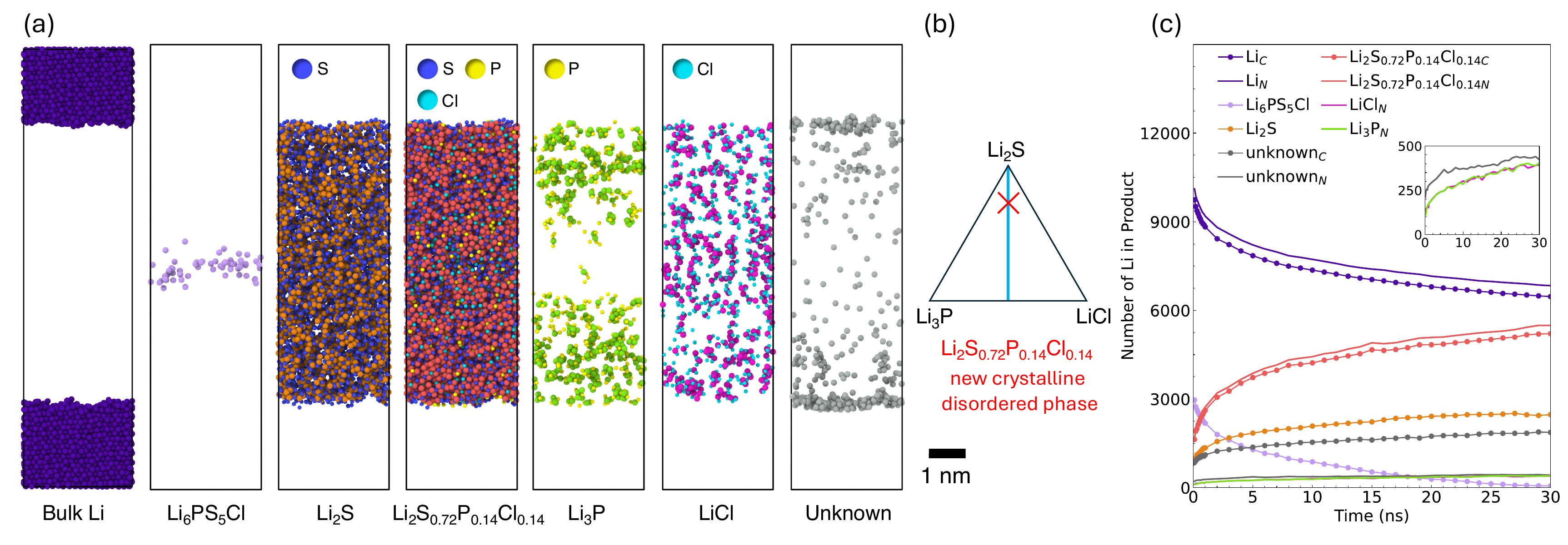}
\caption{\textbf{Unsupervised characterization of interfacial reactions.} (a) Identification of local structures in the reacted interface from MLMD simulations at 30\,ns. Lithium atoms are color-coded according to their identified structural environments. Non-lithium species are shown with colors as indicated in the legend, except in ``unknown" structures where they are omitted for clarity. (b) Phase diagram of {\product}. (c) Time evolution of detected phases at each step of classification. The inset shows a zoomed-in view of the low abundance region, highlighting the presence of LiCl and Li$_3$P from the non-core clustering. Subscripts $C$ and $N$ denote core and non-core clustering}
\label{fig:DPA_clustering} 
\end{figure*}

\subsection*{SEI composition from unsupervised classification:\\ the discovery of a new crystalline disordered phase}
Understanding the formation mechanism of the SEI remains challenging due to its complex, heterogeneous nature. Conventional statistical approaches, such as radial distribution function (RDF) analysis or simple phase identification methods, such as polyhedral template matching, frequently struggle with the amorphous character of evolving reaction zones. Supervised classifiers require prior knowledge of the full set of candidate structures and assign all configurations to classes present in the training set. This limitation renders them unsuitable for detecting and discovering any SEI products beyond those included in the training labels. Unsupervised clustering approaches, instead, are learning techniques that group data by similarity without prior notion of classes. Subsequently, by examining the contents of each cluster, one can assign labels according to common properties among the members. If no common property is identified, the clusters remains unlabeled. This approach, which we refer to as unsupervised classification, is better suited for analyzing complex reactions in which not all products are known a priori, compared to supervised classification methods.

We developed an unsupervised classifier to track the temporal and spatial evolution of reactants and reaction products, specifically capable of identifying the composition and morphology of the SEI and detecting new phases beyond known and expected crystal structures (i.e. labeled data). We used the atomic cluster expansion (ACE) descriptors \cite{drautz2019atomic}, centered on Li atoms, to represent all compounds based on their local structure features. Our SEI characterization workflow, illustrated in Fig.~\ref{fig:DPA_workflow}, involves four key steps: (1) mapping Li-centered atomic environments to high-dimensional ACE descriptors, (2) performing density estimation and clustering in ACE descriptor space based on Euclidean distances between neighboring points, using PA$k$ density estimator \cite{rodriguez2018computing} and density peak advanced (DPA) clustering \cite{dErrico2021automatic}, as implemented in the DADApy package \cite{glielmo2022dadapy}, (3) assigning labels to the resulting clusters, and (4) exploring the contents of ``unknown'' clusters, with potential to augment the set of labeled data. Further details were provided in the Supplementary Information.

Each cluster contains one density peak and multiple saddle points [Fig.~\ref{fig:DPA_workflow} (a)]. Cluster cores (dark shaded colors) are defined as regions where the density value exceeds the highest-density saddle point among the set of adjacent saddle points \cite{Sormani2019,Carli2020CandidateSimulations}. We note that clustering is performed only once, but additional labeled data can be incorporated at any time \cite{Carli2021Interpolation} and assigned to the existing clusters. 
In this approach, core regions correspond to crystalline configurations, starting with the expected structures, while the non-core regions (light shaded colors) represent less frequent geometries, either thermally more disordered or higher-energy configurations.

For the initial labeling of clusters, we generated a reference dataset from MLMD and AIMD trajectories at 300\,K of known crystal structures for the reactants (Li, {\Argy}) and thermodynamically expected products ({\product}). These reference structure sets were then assigned to the clusters already identified in our reacted MLMD simulations. Clusters containing these labeled reference data in their cores were assigned the corresponding label, while clusters with cores not containing any reference structures were classified as ``unknown''. 
In the first pass, the algorithm identified Li$_2$S as the sole crystalline product, alongside the crystalline reactants and significant amounts of ``unknown'' structural environments, as shown in Fig.~S10. Their corresponding abundances as a function of time are shown in Fig.~S11. Interestingly, only a minor portion of the SEI was identified as crystalline Li$_2$S -- for example, 27\% at 30\,ns. The remaining portion, though with a visually obvious crystalline region, remained unknown, suggesting the formation of a previously unreported phase. We referred to this phase as the crystalline disordered phase, Li$_2$S$_{0.72}$P$_{0.14}$Cl$_{0.14}$, which exhibits altered properties compared to Li$_2$S. A detailed discussion of the structures, compositions, and properties -- including its electronic semiconducting and ionic conducting characteristics -- is provided in the next section.

Upon introducing additional labels for this crystalline disordered phase, four previously "unknown" clusters, containing most of the unlabeled structures (82\% at 30\,ns), were classified as this new phase [see Fig.~S10 and S11]. The number of identified Li$_2$S remains unchanged.
Finally, by examining the non-core regions of all remaining ``unknown'' clusters, we detected LiCl and Li$_3$P local environments. However, the appearance of Li$_3$P and LiCl exclusively in non-core regions means that these structures only partially resembled crystalline Li$_3$P and LiCl.

From the spatial distribution of classified local environments in Fig.~\ref{fig:DPA_clustering}, we observed that 
%Li$_2$S and Li$_2$S$_{0.72}$P$_{0.14}$Cl$_{0.14}$ predominantly appeared in the SEI as crystalline domains often located adjacent to each other, and
small domains of Li$_2$S appeared intermixed with the majority product  Li$_2$S$_{0.72}$P$_{0.14}$Cl$_{0.14}$ crystalline disordered phase. This is because some local environments with no neighboring Cl or P are classified as Li$_2$S, given the structural similarity. In contrast, Li$_3$P and LiCl, which would be structurally distinct, appeared mostly as isolated local configurations, with Cl or P ions typically surrounded by Li ions exceeding their stoichiometric ratios. This result explains recent XPS observations \cite{alt2024quantifying}, which showed significant contribution Li$_x$P with only minor bulk-phase Li$_3$P signals. 

The evolution of all reactants and reaction products over time, including both core and non-core clustering, was shown in Fig.~\ref{fig:DPA_clustering}. The observed decrease in Li anode and electrolyte content aligned with the reduction in PS$_4$ tetrahedra previously discussed in Fig.~\ref{fig:Reaction_mechanism} (d). According to our simulation, by 30\,ns, when the entire {\Argy} has been consumed, the SEI is composed of approximately 60\% Li$_2$S$_{0.72}$P$_{0.14}$Cl$_{0.14}$, 27\% crystalline Li$_2$S, 4\% LiCl, 4\% Li$_3$P, and 5\% unknown phase.

\begin{figure*}[ht!]
\includegraphics[width=1\linewidth,trim={0cm 0cm 0cm 0cm},clip]{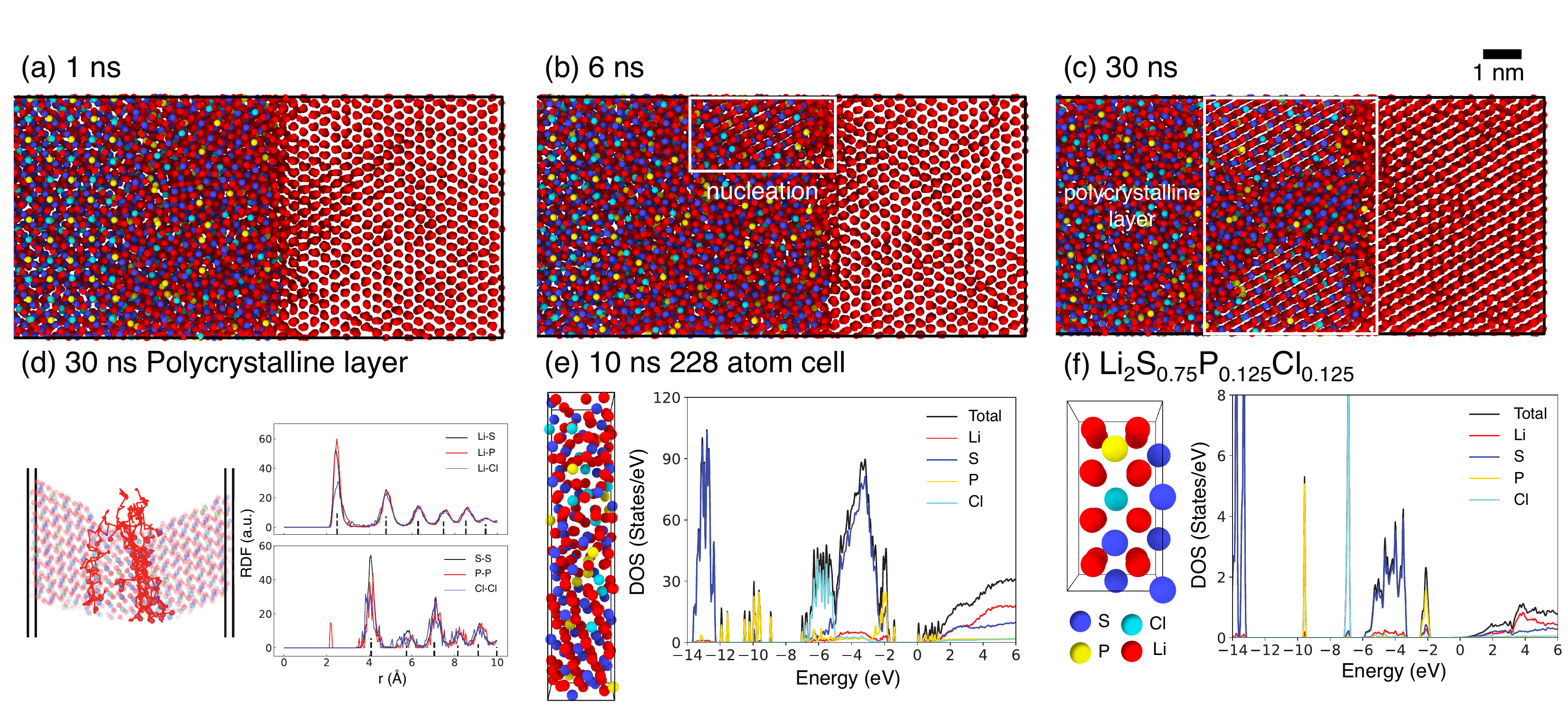}
\caption{\textbf{Time evolution of SEI and physical properties of the crystalline disordered phase}.  
(a-c) Time evolution of a 21,120-atoms structure showing SEI formation. The SEI remains mostly amorphous in the early stages, with nucleation beginning around 5\,ns. A polycrystalline layer form by 30\,ns. 
(d) The polycrystalline region and its radial distribution function (RDF), with black dashed lines indicating theoretic peaks of Li$_2$S crystals. This region contains approximately 14\% P and Cl, with overall stoichiometry and structure matching which of Li$_2$S. Li-ion motion within the SEI is mostly restricted to grain boundaries and defect migrations. Trajectories show all Li ions with displacements exceeding 10\,{\AA} over 5\,ns. Electronic density of states (EDOS) for (e) a fully reacted half cell with 228 atoms from MLMD simulations, and (f) Li$_2$S$_{0.75}$P$_{0.125}$Cl$_{0.125}$. The SEI is found to be semiconducting with a band gap of 1.2\,eV. Fermi energies are aligned with the valence band maximum and set to zero. 
} 
\label{fig:SEI_evolution}
\end{figure*}

%\subsection*{SEI product formation and evolution and passivation mechanism}
\subsection*{SEI formation and passivation mechanism: \\critical role of the new disordered crystalline phase}
Figures~\ref{fig:SEI_evolution} and S8 illustrate the main aspects of the interfacial reaction process: (1) formation of the amorphous phase, (2) nucleation of crystalline phases in the SEI, and (3) formation of a polycrystalline layer. The amorphous product consistently appeared first at the electrolyte-SEI boundary, in correspondence with previous simulation \cite{chaney2024two}. 
As shown in Fig.~S8 (b), initial crystal nucleation cores formed during the early stage around 1\,ns but did not immediately developed into a large single crystal. 
Instead, a polycrystalline layer formed by approximately 30\,ns, at which point the SEI growth slowed down, due to mass transport limitations. 

As a point of comparison, an earlier study of the SEI formation in Li$|$Li$_3$PS$_4$$|$Li \cite{ren2024visualizing}
indicated the presence of numerous point defects, such as Li vacancies, at the early stage of the reaction, while the final stage consisted of a stabilized crystalline Li$_2$S region \cite{ren2024visualizing}. According to experiments, the rate capability of Li$|$Li$_3$PS$_4$$|$Li symmetric cells decayed dramatically due to the aggregation of electrochemically inactive Li$_2$S \cite{conder2017direct,yang2021rich,qi2023electrochemical}. 

For {\Argy}, our MLMD simulation revealed a different path of passivation, influenced by the presence of additional Cl atoms, where the mixture of P$^{3-}$ and Cl$^-$ balanced the charges of S$^{2-}$. To study the resulting polycrystalline layer in detail, it was carved out from the MLMD snapshot at 30\,ns, as shown in \ref{fig:SEI_evolution} (d). The overall lattice geometry closely resembled Li$_2$S, as revealed by the RDF, where the Li-P, Li-Cl distributions overlapped with Li-S, and the P-P, Cl-Cl distributions overlapped with S-S. A significant fraction of S sites was occupied by P and Cl atoms ($\sim$14\%). %This indicates the possibility of formation of a phase: Li$_2$S$_{0.72}$P$_{0.14}$Cl$_{0.14}$. 
 
Conventionally, the formation of SEI between Li metal and {\Argy} was analyzed using the quaternary phase diagram of Li-P-S-Cl, often visualized in a ternary manner with stable binary endpoints: Li$_2$S-P$_2$S$_5$-LiCl. However, the presences of mixed phases such as Li$_2$S$_{0.72}$P$_{0.14}$Cl$_{0.14}$, was not predicted by the thermodynamic approach, and this structure does not appear in any crystal database. We further discuss in the Discussion section why this dominant phase remained undetected in XPS and XRD experiments. The fact that this phase rapidly nucleated and grew in our MLMD simulations indicats the significant role in the SEI formation process that is played by diffusion and nucleation kinetics, in competition with thermodynamics. We closely examined this newly identified phase, considering its thermodynamic stability and electronic structure.

To assess the formation stability of the crystalline disordered phase, we calculated the DFT formation energy for all symmetry-inequivalent structures of Li$_2$S$_{0.75}$P$_{0.125}$Cl$_{0.125}$ (24 atoms), which had a concentration close to Li$_2$S$_{0.72}$P$_{0.14}$Cl$_{0.14}$. The formation energy of the most stable structure was found to be 27.7\,meV/atom above the convex hull, within the expected range for thermodynamically stable or meta-stable crystals \cite{sun2016thermodynamic,hegde2020phase}, considering also the average error of the PBE DFT functional. In addition, the S-P-Cl disorder is further stabilized by configurational entropy, which in the random solution amounts to approximately 20 \,meV/atom at room temperature. Therefore, we conclude that the newly identified crystalline disordered phase has a high probability of existence in Argyrodite-based ASSBs due to favorable nucleation and growth kinetics, as well as accessible thermodynamic (meta)stability aided by disorder.

For a product phase in the SEI to act as a passivation layer, it must impede the transport of electrons or ions, or both, since both are needed in order for the redox reaction to proceed indefinitely.
To assess electrochemical stability and passivation characteristics, we calculated the electronic density of states using DFT. We considered three scenarios, bare Li$_2$S (8 atoms), Li$_2$S with low concentration doping: Li$_2$S$_{0.9375}$P$_{0.03125}$Cl$_{0.03125}$ (32 atoms), and Li$_2$S$_{0.75}$P$_{0.125}$Cl$_{0.125}$ (24 atoms), as shown in Fig.~S14. The disordered site configurations were randomly selected for Li$_2$S$_{0.9375}$P$_{0.03125}$Cl$_{0.03125}$. All four symmetry-inequivalent structures of Li$_2$S$_{0.75}$P$_{0.125}$Cl$_{0.125}$ were included. For Li$_2$S, we obtained a band gap of 3.4\,eV using the PBE functional. Introducing P and Cl resulted in an additional defect band in the original band gap. The Fermi energy is defined at the conduction band minimum. For all seven representative structures, we found the band structures to be semiconducting (Fig.~S14) in DFT with band gap values ranging from 1.4 to 2.0\,eV. Thus, we concluded that the crystalline disordered phase can effectively suppress the electron transport, thus passivating the interfacial reaction. To characterize whether this crystalline disordered phase affected the ionic diffusivity, we extended the MLMD simulation to 35\,ns and plotted the Li ion trajectories with the displacement amplitudes exceeding 10\,{\AA} over 5\,ns, shown as the red solid lines in Fig.~\ref{fig:SEI_evolution} (d). Li-ion diffusion was clearly dominated by grain boundaries between the two crystalline domains. Some defect migration also occurred.
To study bulk diffusion in the new phase we performing MLMD on the supercells of the representative Li$_2$S$_{0.75}$P$_{0.125}$Cl$_{0.125}$ (1536 atoms) structures. From the time-averaged MSD, some disordered configurations exhibited diffusion coefficients of up to 4.2$\times$10$^{-9}$\,cm$^2$s$^{-1}$ (Fig.~S15). For compounds with exact stoichiometry, Li ion diffusion occurred through defects in the crystal, where a vacant site for a subsequent jump was available when a Li ion jumped into an interstitial site. Hence, this new phase is moderately ionically conducting, which allows the reaction to proceed even after its formation.

\begin{figure}[htbp]
\includegraphics[width=1\linewidth,trim={0cm 0cm 0cm 0cm},clip]{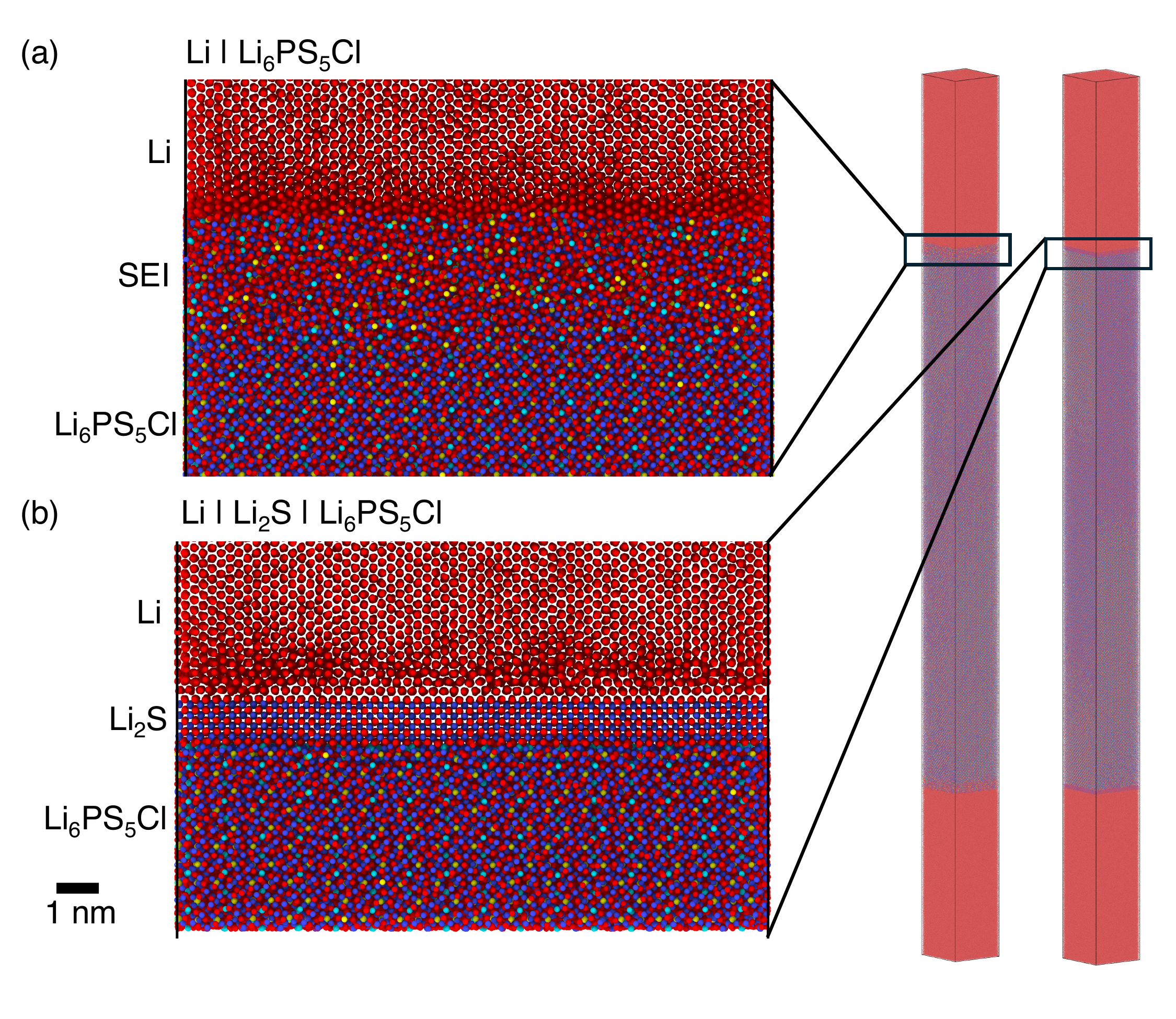}
\caption{\textbf{Reactive simulations of three-million-atom cells demonstrating the robustness of the MLIP.} (a) 3,412,960-atom cell simulation around 1\,ns, where the uncoated Li$|${\Argy} reacted similarly to smaller cells. (b) 3,450,592-atom cell simulation with two layers of Li$_2$S  inserted between Li and {\Argy}, showing no reaction after 1\,ns. (c) Full cells of Li$|${\Argy}$|$Li and Li$|$Li$_2$S$|${\Argy}$|$Li$_2$S$|$Li.
} 
\label{fig:Li2S_protection}
\end{figure}

In comparison with the disordered phase, ideal crystalline Li$_2$S behaves differently in terms of transport characteristics, dramatically altering the interface reaction. We explicitly introduced very thin layers of Li$_2$S between Li and {\Argy} and performed simulations with over three million atoms for up to 1\,ns at 300\,K, as illustrated in Fig.~\ref{fig:Li2S_protection}. 
The uncoated system exhibited rapid reactions similar to those reported above for the smaller cells. In contrast, the Li$_2$S coating, with only two unit cell layers, completely prevented reactions between Li and {\Argy} from even initiating in the course of the entire simulation, as there was no observable Li-ion transport through the pure Li$_2$S layer. On the other hand, the ability of Li$_2$S$_{0.72}$P$_{0.14}$Cl$_{0.14}$ to conduct Li-ion through the interstitial mechanism mentioned above allows the battery to operate and the SEI to grow.

\section*{Discussion} 
Our large-scale simulations provide direct and unbiased observations focusing on the onset and passivation of the interfacial reactions between anode Li and {\Argy}, corroborating experimental hypotheses and enhancing confidence in predicting SEI evolution. 

Our finding is that the interplay of fast ion interdiffusion with significant thermodynamic driving forces for reactions leads to the formation of a kinetically stabilized product that was not thermodynamically expected or  known to exist. The SEI between Li and {\Argy} comprises a highly modified crystalline phase, with a representative stoichiometry of  Li$_2$S$_{0.72}$P$_{0.14}$Cl$_{0.14}$, which is structurally similar to the thermodynamically expected Li$_2$S but, unlike Li$_2$S, has moderate Li$^+$ conductivity and allows the reaction to proceed as long as ions can migrate to form additional products. It is also a semiconductor that can effectively passivate electron transport, but with a smaller band gap than Li$_2$S.
Why did this crystalline disordered phase remain hidden? Crucially, current computational approaches only include known stable phases. Discovery of new phases is only made possible by unbiased explicit reactive MD, based on first principles, together with unsupervised classification.
Experimental studies of the reaction between Li and {\Argy} commonly identified the dominant composition of the crystalline layer as Li$_2$S from XRD or XPS \cite{pang2017vivo,wenzel2018interfacial,tan2019elucidating,narayanan2022effect,otto2022situ,aktekin2023sei,alt2024quantifying}. Li$_3$P on the other hand, was often detected with low intensities due to possible amorphous structures, such as the presence of various lithiated phosphorous compounds(Li$_x$P), and high reactivity with oxygen and moisture \cite{wenzel2018interfacial,narayanan2022effect,aktekin2023sei,alt2024quantifying}. LiCl could not be detected by XPS due to the negligible binding energy difference between Cl$^-$ in LiCl and {\Argy}. Its presence, however, was suggested by XRD in the directly synthesized SEI \cite{alt2024quantifying}. 
For XPS, the overall similarity in the structure of Li$_2$S$_{0.72}$P$_{0.14}$Cl$_{0.14}$ and Li$_2$S results in similar core-level photoemission spectra for S $2p$. P $2p$ and Cl $2p$ signals are likely hidden due to the reasons mentioned above.
As for XRD, we extracted all crystalline disordered phases from the SEI using our unsupervised classifier and calculated its XRD pattern. Interestingly, the pattern closely resembles the ideal Li$_2$S in both peak positions and relative intensities Fig.~S12, despite broader features due to thermal vibration, crystal size, defects and microstrain etc. This suggested possible overlook in experiments, such as XRD, based on thermodynamically stable products. Due to the mixed nature of multiple phases in the SEI, verifying its electronic or ionic properties would also be challenging. One way to verify the existence of this new phase could be through direct synthesis with stoichiometry suggested by our MLMD simulations.

Despite efforts to enhance the accuracy of the MLIP, inherent limitations still exist that require attention in future research. At the DFT level, semi-local functionals such as PBE often struggle to accurately represent electron localization and charge transfer processes. Ground state DFT charge distributions do not capture the physics of electron transport, which is important when considering the role of insulating SEI products. Although an insulating layer should impede electron transport, spurious electron tunneling may still occur in DFT simulations, and may facilitate reactions in DFT and MLIP simulations that in reality would be suppressed by passivation. Generally, describing kinetics of redox and electron transfer reactions in atomistic simulations remains an open challenge. We note that the presented approach would be more reliable for simulating and analyzing solid-state reactions that do not involve changes of oxidation state, where electronic transport is not relevant.
An additional consideration is that long-range electrostatic interactions may become important in simulations of heterogeneous systems.
Our Allegro MLIP model incorporates electrostatic interactions only up to the prespecified cutoff radius. By varying the cutoff radius, we did not observe qualitatively different accuracies during model training, which indicates that this is not a major concern.

We have resolved the detailed mechanisms of the onset of the interfacial reactions between anode Li and {\Argy}. However, to reach the final passivation, much longer simulations are needed to establish the exact nature of the transport limitations in the later stages of the SEI growth. Furthermore, the reaction rate and mechanical effects depend  on the simulation cell size and composition. We examined the size effect by varying the number of atoms and the initial fraction of Li metal relative to the amount of {\Argy}. The SEI thickness as a function of time is consistent in the three cases with 21,120, 120,960 and 3,412,960 atoms (see Fig.~S17). 
However, a smaller cell containing 10,080 atoms (in-plane dimensions of 2\,nm$\times$3\,nm versus 6\,nm$\times$3\,nm for the 21,120-atom cell) leads to a faster single crystalline layer formed in as little as 200\,ps (see Fig~S18). A simulation cell large enough in directions perpendicular to the growth direction is needed to converge the SEI formation kinetics.

We have demonstrated the power of explicit first-principles MLMD as a digital twin, capable of elucidating complex kinetic and mechanical interfacial phenomena in batteries and in solid state reactions generally. Integration on high-fidelity MLIPs developed through on-the-fly Bayesian active learning with advanced unsupervised clustering techniques is shown to be a promising path for accelerating and automating the simulations and analysis of the SEI formation at the atomic level. 

The interfacial process considered here is an example of a solid state chemical reaction, and so the development is also laying the foundation for the understanding of the solid-state synthesis mechanism. Given the lack of quantitative general methods for predicting realistic solid interfacial reactions, including contributions of thermodynamics and kinetics, large-scale direct reactive MLMD, with its first principles accuracy and scalability, is a promising approach for exploring interfacial engineering.
The combined digital-twin approach explicitly considered SEI composition, structural evolution, and spatial distribution of reaction products at the atomic level, without any assumptions or fits to experiment, relying only on controlled approximations starting from first principles. By providing explicit atomistic details, this work goes beyond previous homogenized continuum models of solid state reactions \cite{chromik1999thermodynamic,perez2006combined,gusak2019phase,chen2022classical}.
In particular, unsupervised classification of atomic environments is capable of capturing complex reaction product geometries, identifying previously unknown phases of matter, and adapting to the structural diversity generated from MD simulations.
The ability to track the evolution of phases in time and space at the atomic level has allowed us to reveal the importance of kinetic factors in governing SEI evolution, challenging conventional thermodynamic predictions. The presence of complex, crystalline disordered phase within the SEI, which are not captured by traditional phase diagrams, suggests that kinetics are more important than previously thought in governing interface reactions in solid-state batteries, due to the highly mobile nature of Li ions. These findings highlight the necessity of employing large-scale reactive MLMD simulations to capture the dynamic and heterogeneous nature of these interfaces, which are critical for optimizing the performance and stability of advanced energy storage systems.

\section*{Acknowledgments}
The authors acknowledge the fruitful discussion with Zachary A. H. Goodwin and Julia H. Yang and the software support from Anders Johansson, Simon Batzner. Model development, MLMD simulation, and data analysis by J.D. was supported by the National Science Foundation under Grant No. DMR-2119351. Unsupervised classifier development by L.Z. M.C. was supported by the Department of Navy award N00014-20-1-2418 issued by the Office of Naval Research. Data analysis by M.W. was supported by the National Science Foundation Harnessing the Data Revolution Big Idea under Grant No. 2118201. A.M. was supported by U.S. Department of Energy, Office of Science, Office of Advanced Scientific Computing Research, Computational Science Graduate Fellowship under Award Number(s) DESC0021110. Y.X. was supported by the US Department of Energy (DOE) Office of Basic Energy Sciences under Award No. DE-SC0020128. B.K. acknowledges support from Robert Bosch LLC.
An award for computer time was provided by the U.S. Department of Energy’s (DOE) Innovative and Novel Computational Impact on Theory and Experiment (INCITE) Program.
This research used resources of the Argonne Leadership Computing Facility, which is a DOE Office of Science User Facility supported under Contract DE-AC02-06CH11357, the Delta advanced computing and data resource which is supported by the National Science Foundation (award OAC 2005572) and the State of Illinois and the Oak Ridge Leadership Computing Facility at the Oak Ridge National Laboratory, which is supported by the Office of Science of the U.S. Department of Energy under Contract No. DE-AC05-00OR22725.
Additional computing resources were provided by the Harvard University FAS Division of Science Research Computing Group.

\section*{Author contributions}
B.K. and J.D. designed the project. J.D. generated the datasets, trained and validated MLIPs, conducted MLMD simulations. J.D. and M.W. performed analysis of MLMD simulations and initiated atomic environment based classification. L.Z. and M.C. designed and implemented unsupervised approach of classification and performed further analysis. A.M. provided support for Allegro implementation and model training. Y.X. provided support for FLARE implementation and training data generation. All authors contributed to the manuscript.

\section*{Competing interests}
The authors declare no competing interests.

\section*{Data and materials availability} 
All data needed to evaluate the conclusions in the paper are present in the paper and/or the Supplementary Materials. Additional data related to this paper may be requested from the authors. The numerical data for the figures are available from the Harvard Dataverse Repository at https://doi.org/10.7910/DVN/NUHF3O. \textbf{Code availability:} The codes that support findings of the study are available from the corresponding authors on request.

\section*{Methods}
\subsection*{Model training procedure}
The active learning from Flare \cite{vandermause2020fly,vandermause2022active,xie2023uncertainty} allowed the collection of uncorrelated frames up to 1\,ns molecular dynamics (MD) trajectory at each temperature and each pressure with only a few tens to a few hundred DFT calculations. We performed on-the-fly active learning on Li, {\Argy}, and {\symcell}. The temperatures were 100 to 800\,K with 100\,K increment, and at 2,000\,K to sample a molten phase. 1, 5, and 10\,GPa and 1\%, 5\%, and 10\% lattice expansion were also considered at 300\,K. 1492, 816, and 2106 frames were collected for Li, {\Argy}, and {\symcell}, respectively.
During the active learning workflow, we used PyLAMMPS MD engine with 1\,fs timestep. The structures were jittered by 0.01\,{\AA} to create the initial Bayesian force field through sparse Gaussian Process (SGP). Hyperparameters were updated starting from the second frame. We used the B2 descriptor with 6\,{\AA} radial cutoff, $n_{max}$ = 9, $l_{max}$ = 3. The active learning workflow used model to predict uncertainty and to decide whether to keep exploring the configuration space, or call DFT to collect new training data. For BCC Li, the global uncertain threshold to call DFT was 0.0005 (maximum of all atomic environments), and the atomic environments with uncertainty greater than 0.0001 were added to the SGP. These very small values were selected because the BCC Li had small diversity between atomic environments. For {\Argy} and {\symcell}, the global uncertain threshold was 0.15, and atomic environment threshold was 0.03. The MD exploration was done with NPT ensemble.

The strictly local equivariant neural network model Allegro models \cite{musaelian2023learning} used two layers of 8 tensor features. We used a radial cutoff of 6\,{\AA}, and Bessel basis 8 and polynomial cutoff 6 for radial basis, $l_{max}$ = 2, full $O$(3) symmetry. The two-body latent multi-layer perceptron (MLP) hidden dimensions were [32, 64, 128] and [128, 128] respectively, both with SiLU nonlinearities. The final edge energy had a single linear hidden layer of dimenstion 32. Per-species shifts were calculated from the single atom energy and were applied due to the complicated atomic environments, e.g., Li appeared in BCC Li, {\Argy}, {\symcell} and reaction products. The values for Li, S, P, and Cl were -0.29767466, -1.07917909, -1.88667948, and -0.2581321, respectively. We used all energy, force, and stress labels for the training, and we weighted them as 10, 1, 100 and trained on per atom MSE loss. The training and validation sets were 90\% and 10\%, 7325, and 814 frames. Production MLMD frames recalculated with DFT were used for the test set. We use a batch size of 5, a maximum epochs 100,000, and learning rate 0.002. The data set was re-shuffled after each epoch. A Ziegler-Biersack-Littmark (ZBL) potential were used to enhance the stability of the model \cite{ziegler1985stopping}.

To improve the fidelity of the Allegro model, additional data from the production run of all the three systems from an initial Allegro model, and from \textit{ab initio} molecular dynamics (AIMD) of thermodynamically proposed reaction products LiCl, Li$_2$S, Li$_3$P \cite{zhu2015origin} were collected. Two iterations of training and MD augmentation were performed to get the final reactive model. Bulk properties were captured by training on solo composition, and the interfacial reaction kinetics was represented by a {\symcell} with 228 atoms. Nonphysical cavity appearing in the {\symcell} from the initial model was removed from the final model. The final reactive model achieved excellent accuracy on all the benchmarks. The detailed list for the training were summarized in Table.~S1.

\subsection*{DFT, MD parameters}
Theoretical calculations were performed with the Vienna \textit{Ab initio} Simulation Package (VASP 6) \cite{Kresse1993,KresseFurthPRB,KresseFurthCMS}, using the generalized gradient approximation (GGA) Perdew-Burke-Erzenhof (PBE) exchange-correlation functional \cite{refLDA,refPBE}. We used projected-augmented-wave potentials, which explicitly includes 3 valence electrons for Li (1S$^2$2S$^1$), 6 for S (3S$^2$3P$^4$), 5 for P (3S$^2$3P$^3$), and 7 for Cl (3S$^2$3P$^5$). We relaxed the BCC Li lattice parameter with a $k$-point mesh of 15$\times$15$\times$15 with a plane-wave energy cut-off of 700\,eV and the electronic self-consistent loop convergence of 10$^{-8}$\,eV. The structure were relaxed until the forces being smaller than 10$^{-5}$\,eV/{\AA}, yielding a lattice parameter of 3.420\,{\AA}, less than 0.5\% difference compared to previous DFT simulations \cite{zuo2020performance,phuthi2024accurate}. For FLARE on-the-fly (OTF) data collection, electronic $k$-point mesh varied with the structures. We used 3$\times$3$\times$3 for 3$\times$3$\times$3 supercell of Li (54 atoms), 4$\times$4$\times$4 for unitcell of {\Argy} (52 atoms), and 1$\times$1$\times$1 for {\symcell} (228 atoms). All $k$-point meshes centered at $\Gamma$-point. All energies, forces, and stress are collected for the model train.

AIMD simulations were performed using \textit{NVT} with a Nos$\acute{e}$--Hoover thermostat. MLMD simulations were performed with LAMMPS \cite{LAMMPS} using \textit{NPT} with Nos$\acute{e}$--Hoover thermostat and barostat with time constant 0.1\,ps and 1\,ps. The initial configuration of {\Argy} was determined where 24 Li$^+$ were randomly removed from the 48$h$ sites to ensure the stoichiometry of the 50\% occupancy. 50\% disorder of Cl$^-$/S$^{2-}$ was also randomly generated. 

The Li to {\Argy} ratio varied across the three configurations, influenced both by design and the need to minimize lattice mismatch.
In the 21,120 cell, we doubled the amount of anode Li (16:1) to consume more electrolyte, compared to the thermodynamically predicted pathway $Li_6PS_5Cl + 8Li \rightleftharpoons 5Li_2S + Li_3P + LiCl$. The 120,960 cell had the same Li to {\Argy} ratio (8:1) as the chemical equation. In the two three-million-atom cells, the Li and {Argy} approximated 8:1.
In all the three configurations, to reduce lattice mismatch, {\Argy} (100) plane was aligned with Li (110) plane, resulting in -0.4\% and -1.4\% mismatch along Li [111], and Li [11-2] directions, respectively.

\newpage

\bibliography{References}

\begin{thebibliography}{10}

\bibitem{janek2016solid}
J{\"u}rgen Janek and Wolfgang~G Zeier.
\newblock A solid future for battery development.
\newblock {\em Nature energy}, 1(9):1--4, 2016.

\bibitem{liu2019pathways}
Jun Liu, Zhenan Bao, Yi~Cui, Eric~J Dufek, John~B Goodenough, Peter Khalifah,
  Qiuyan Li, Bor~Yann Liaw, Ping Liu, Arumugam Manthiram, et~al.
\newblock Pathways for practical high-energy long-cycling lithium metal
  batteries.
\newblock {\em Nature Energy}, 4(3):180--186, 2019.

\bibitem{kamaya2011lithium}
Noriaki Kamaya, Kenji Homma, Yuichiro Yamakawa, Masaaki Hirayama, Ryoji Kanno,
  Masao Yonemura, Takashi Kamiyama, Yuki Kato, Shigenori Hama, Koji Kawamoto,
  et~al.
\newblock A lithium superionic conductor.
\newblock {\em Nature materials}, 10(9):682--686, 2011.

\bibitem{kato2016high}
Yuki Kato, Satoshi Hori, Toshiya Saito, Kota Suzuki, Masaaki Hirayama, Akio
  Mitsui, Masao Yonemura, Hideki Iba, and Ryoji Kanno.
\newblock High-power all-solid-state batteries using sulfide superionic
  conductors.
\newblock {\em Nature Energy}, 1(4):1--7, 2016.

\bibitem{adeli2019boosting}
Parvin Adeli, J~David Bazak, Kern~Ho Park, Ivan Kochetkov, Ashfia Huq,
  Gillian~R Goward, and Linda~F Nazar.
\newblock Boosting solid-state diffusivity and conductivity in lithium
  superionic argyrodites by halide substitution.
\newblock {\em Angewandte Chemie}, 131(26):8773--8778, 2019.

\bibitem{yu2016synthesis}
Chuang Yu, Lambert van Eijck, Swapna Ganapathy, and Marnix Wagemaker.
\newblock Synthesis, structure and electrochemical performance of the
  argyrodite li6ps5cl solid electrolyte for li-ion solid state batteries.
\newblock {\em Electrochimica Acta}, 215:93--99, 2016.

\bibitem{zheng2018review}
Feng Zheng, Masashi Kotobuki, Shufeng Song, Man~On Lai, and Li~Lu.
\newblock Review on solid electrolytes for all-solid-state lithium-ion
  batteries.
\newblock {\em Journal of Power Sources}, 389:198--213, 2018.

\bibitem{famprikis2019fundamentals}
Theodosios Famprikis, Pieremanuele Canepa, James~A Dawson, M~Saiful Islam, and
  Christian Masquelier.
\newblock Fundamentals of inorganic solid-state electrolytes for batteries.
\newblock {\em Nature materials}, 18(12):1278--1291, 2019.

\bibitem{banerjee2020interfaces}
Abhik Banerjee, Xuefeng Wang, Chengcheng Fang, Erik~A Wu, and Ying~Shirley
  Meng.
\newblock Interfaces and interphases in all-solid-state batteries with
  inorganic solid electrolytes.
\newblock {\em Chemical reviews}, 120(14):6878--6933, 2020.

\bibitem{zhang2023homogeneous}
Qian-Kui Zhang, Xue-Qiang Zhang, Jing Wan, Nan Yao, Ting-Lu Song, Jin Xie,
  Li-Peng Hou, Ming-Yue Zhou, Xiang Chen, Bo-Quan Li, et~al.
\newblock Homogeneous and mechanically stable solid--electrolyte interphase
  enabled by trioxane-modulated electrolytes for lithium metal batteries.
\newblock {\em Nature Energy}, 8(7):725--735, 2023.

\bibitem{schwietert2020clarifying}
Tammo~K Schwietert, Violetta~A Arszelewska, Chao Wang, Chuang Yu, Alexandros
  Vasileiadis, Niek~JJ de~Klerk, Jart Hageman, Thomas Hupfer, Ingo Kerkamm,
  Yaolin Xu, et~al.
\newblock Clarifying the relationship between redox activity and
  electrochemical stability in solid electrolytes.
\newblock {\em Nature materials}, 19(4):428--435, 2020.

\bibitem{kasemchainan2019critical}
Jitti Kasemchainan, Stefanie Zekoll, Dominic Spencer~Jolly, Ziyang Ning,
  Gareth~O Hartley, James Marrow, and Peter~G Bruce.
\newblock Critical stripping current leads to dendrite formation on plating in
  lithium anode solid electrolyte cells.
\newblock {\em Nature materials}, 18(10):1105--1111, 2019.

\bibitem{ning2023dendrite}
Ziyang Ning, Guanchen Li, Dominic~LR Melvin, Yang Chen, Junfu Bu, Dominic
  Spencer-Jolly, Junliang Liu, Bingkun Hu, Xiangwen Gao, Johann Perera, et~al.
\newblock Dendrite initiation and propagation in lithium metal solid-state
  batteries.
\newblock {\em Nature}, 618(7964):287--293, 2023.

\bibitem{fang2019key}
Chengcheng Fang, Xuefeng Wang, and Ying~Shirley Meng.
\newblock Key issues hindering a practical lithium-metal anode.
\newblock {\em Trends in Chemistry}, 1(2):152--158, 2019.

\bibitem{chen2021new}
Xiao-Ru Chen, Chong Yan, Jun-Fan Ding, Hong-Jie Peng, and Qiang Zhang.
\newblock New insights into “dead lithium” during stripping in lithium
  metal batteries.
\newblock {\em Journal of Energy Chemistry}, 62:289--294, 2021.

\bibitem{jin2021rejuvenating}
Chengbin Jin, Tiefeng Liu, Ouwei Sheng, Matthew Li, Tongchao Liu, Yifei Yuan,
  Jianwei Nai, Zhijin Ju, Wenkui Zhang, Yujing Liu, et~al.
\newblock Rejuvenating dead lithium supply in lithium metal anodes by iodine
  redox.
\newblock {\em Nature Energy}, 6(4):378--387, 2021.

\bibitem{zhu2015origin}
Yizhou Zhu, Xingfeng He, and Yifei Mo.
\newblock Origin of outstanding stability in the lithium solid electrolyte
  materials: insights from thermodynamic analyses based on first-principles
  calculations.
\newblock {\em ACS applied materials \& interfaces}, 7(42):23685--23693, 2015.

\bibitem{richards2016interface}
William~D Richards, Lincoln~J Miara, Yan Wang, Jae~Chul Kim, and Gerbrand
  Ceder.
\newblock Interface stability in solid-state batteries.
\newblock {\em Chemistry of Materials}, 28(1):266--273, 2016.

\bibitem{tan2019elucidating}
Darren~HS Tan, Erik~A Wu, Han Nguyen, Zheng Chen, Maxwell~AT Marple, Jean-Marie
  Doux, Xuefeng Wang, Hedi Yang, Abhik Banerjee, and Ying~Shirley Meng.
\newblock Elucidating reversible electrochemical redox of li6ps5cl solid
  electrolyte.
\newblock {\em ACS Energy Letters}, 4(10):2418--2427, 2019.

\bibitem{chen2019approaching}
Rusong Chen, Qinghao Li, Xiqian Yu, Liquan Chen, and Hong Li.
\newblock Approaching practically accessible solid-state batteries: stability
  issues related to solid electrolytes and interfaces.
\newblock {\em Chemical reviews}, 120(14):6820--6877, 2019.

\bibitem{xiao2020understanding}
Yihan Xiao, Yan Wang, Shou-Hang Bo, Jae~Chul Kim, Lincoln~J Miara, and Gerbrand
  Ceder.
\newblock Understanding interface stability in solid-state batteries.
\newblock {\em Nature Reviews Materials}, 5(2):105--126, 2020.

\bibitem{schmalzried1974solid}
H~Schmalzried.
\newblock Solid-state reactions between oxides: Historical introduction.
\newblock In {\em Defects and transport in oxides}, pages 83--107. Springer,
  1974.

\bibitem{ye2021dynamic}
Luhan Ye and Xin Li.
\newblock A dynamic stability design strategy for lithium metal solid state
  batteries.
\newblock {\em Nature}, 593(7858):218--222, 2021.

\bibitem{ye2024fast}
Luhan Ye, Yang Lu, Yichao Wang, Jianyuan Li, and Xin Li.
\newblock Fast cycling of lithium metal in solid-state batteries by
  constriction-susceptible anode materials.
\newblock {\em Nature materials}, 23(2):244--251, 2024.

\bibitem{pang2017vivo}
Quan Pang, Xiao Liang, Abhinandan Shyamsunder, and Linda~F Nazar.
\newblock An in vivo formed solid electrolyte surface layer enables stable
  plating of li metal.
\newblock {\em Joule}, 1(4):871--886, 2017.

\bibitem{wenzel2018interfacial}
Sebastian Wenzel, Stefan~J Sedlmaier, Christian Dietrich, Wolfgang~G Zeier, and
  J{\"u}ergen Janek.
\newblock Interfacial reactivity and interphase growth of argyrodite solid
  electrolytes at lithium metal electrodes.
\newblock {\em Solid State Ionics}, 318:102--112, 2018.

\bibitem{narayanan2022effect}
Sudarshan Narayanan, Ulderico Ulissi, Joshua~S Gibson, Yvonne~A Chart, Robert~S
  Weatherup, and Mauro Pasta.
\newblock Effect of current density on the solid electrolyte interphase
  formation at the lithium li6ps5cl interface.
\newblock {\em Nature Communications}, 13(1):7237, 2022.

\bibitem{otto2022situ}
Svenja-K Otto, Luise~M Riegger, Till Fuchs, Sven Kayser, Pascal Schweitzer,
  Simon Burkhardt, Anja Henss, and J{\"u}rgen Janek.
\newblock In situ investigation of lithium metal--solid electrolyte anode
  interfaces with tof-sims.
\newblock {\em Advanced Materials Interfaces}, 9(13):2102387, 2022.

\bibitem{alt2024quantifying}
Christoph~D Alt, Nadia~UCB M{\"u}ller, Luise~M Riegger, Burak Aktekin, Philip
  Minnmann, Klaus Peppler, and J{\"u}rgen Janek.
\newblock Quantifying multiphase sei growth in sulfide solid electrolytes.
\newblock {\em Joule}, 8(10):2755--2776, 2024.

\bibitem{cheng2017quantum}
Tao Cheng, Boris~V Merinov, Sergey Morozov, and William~A Goddard~III.
\newblock Quantum mechanics reactive dynamics study of solid
  li-electrode/li6ps5cl-electrolyte interface.
\newblock {\em ACS Energy Letters}, 2(6):1454--1459, 2017.

\bibitem{golov2021molecular}
Andrey Golov and Javier Carrasco.
\newblock Molecular-level insight into the interfacial reactivity and ionic
  conductivity of a li-argyrodite li6ps5cl solid electrolyte at bare and coated
  li-metal anodes.
\newblock {\em ACS Applied Materials \& Interfaces}, 13(36):43734--43745, 2021.

\bibitem{luo2022nanostructure}
Shuting Luo, Xinyu Liu, Xiao Zhang, Xuefeng Wang, Zhenyu Wang, Yufeng Zhang,
  Haidong Wang, Weigang Ma, Lingyun Zhu, and Xing Zhang.
\newblock Nanostructure of the interphase layer between a single li dendrite
  and sulfide electrolyte in all-solid-state li batteries.
\newblock {\em ACS Energy Letters}, 7(9):3064--3071, 2022.

\bibitem{golov2023unveiling}
Andrey Golov and Javier Carrasco.
\newblock Unveiling solid electrolyte interphase formation at the molecular
  level: Computational insights into bare li-metal anode and li6ps5--x se x cl
  argyrodite solid electrolyte.
\newblock {\em ACS Energy Letters}, 8(10):4129--4135, 2023.

\bibitem{hao2024first}
Wei Hao, Swastik Basu, and Gyeong~S Hwang.
\newblock First-principles prediction of reaction-induced structural evolution
  at the interface between lithium metal and sulfide electrolytes.
\newblock {\em The Journal of Physical Chemistry C}, 128(11):4440--4447, 2024.

\bibitem{van2001reaxff}
Adri~CT Van~Duin, Siddharth Dasgupta, Francois Lorant, and William~A Goddard.
\newblock Reaxff: a reactive force field for hydrocarbons.
\newblock {\em The Journal of Physical Chemistry A}, 105(41):9396--9409, 2001.

\bibitem{senftle2016reaxff}
Thomas~P Senftle, Sungwook Hong, Md~Mahbubul Islam, Sudhir~B Kylasa, Yuanxia
  Zheng, Yun~Kyung Shin, Chad Junkermeier, Roman Engel-Herbert, Michael~J
  Janik, Hasan~Metin Aktulga, et~al.
\newblock The reaxff reactive force-field: development, applications and future
  directions.
\newblock {\em npj Computational Materials}, 2(1):1--14, 2016.

\bibitem{behler2007generalized}
J{\"o}rg Behler and Michele Parrinello.
\newblock Generalized neural-network representation of high-dimensional
  potential-energy surfaces.
\newblock {\em Physical review letters}, 98(14):146401, 2007.

\bibitem{bartok2010gaussian}
Albert~P Bart{\'o}k, Mike~C Payne, Risi Kondor, and G{\'a}bor Cs{\'a}nyi.
\newblock Gaussian approximation potentials: The accuracy of quantum mechanics,
  without the electrons.
\newblock {\em Physical review letters}, 104(13):136403, 2010.

\bibitem{shapeev2016moment}
Alexander~V Shapeev.
\newblock Moment tensor potentials: A class of systematically improvable
  interatomic potentials.
\newblock {\em Multiscale Modeling \& Simulation}, 14(3):1153--1173, 2016.

\bibitem{wang2018deepmd}
Han Wang, Linfeng Zhang, Jiequn Han, and E~Weinan.
\newblock Deepmd-kit: A deep learning package for many-body potential energy
  representation and molecular dynamics.
\newblock {\em Computer Physics Communications}, 228:178--184, 2018.

\bibitem{xie2018crystal}
Tian Xie and Jeffrey~C Grossman.
\newblock Crystal graph convolutional neural networks for an accurate and
  interpretable prediction of material properties.
\newblock {\em Physical review letters}, 120(14):145301, 2018.

\bibitem{chen2019graph}
Chi Chen, Weike Ye, Yunxing Zuo, Chen Zheng, and Shyue~Ping Ong.
\newblock Graph networks as a universal machine learning framework for
  molecules and crystals.
\newblock {\em Chemistry of Materials}, 31(9):3564--3572, 2019.

\bibitem{batzner2023}
Simon Batzner, Albert Musaelian, Lixin Sun, Mario Geiger, Jonathan~P Mailoa,
  Mordechai Kornbluth, Nicola Molinari, Tess~E Smidt, and Boris Kozinsky.
\newblock E (3)-equivariant graph neural networks for data-efficient and
  accurate interatomic potentials.
\newblock {\em Nature communications}, 13(1):2453, 2022.

\bibitem{musaelian2023learning}
Albert Musaelian, Simon Batzner, Anders Johansson, Lixin Sun, Cameron~J Owen,
  Mordechai Kornbluth, and Boris Kozinsky.
\newblock Learning local equivariant representations for large-scale atomistic
  dynamics.
\newblock {\em Nature Communications}, 14(1):579, 2023.

\bibitem{vandermause2020fly}
Jonathan Vandermause, Steven~B Torrisi, Simon Batzner, Yu~Xie, Lixin Sun,
  Alexie~M Kolpak, and Boris Kozinsky.
\newblock On-the-fly active learning of interpretable bayesian force fields for
  atomistic rare events.
\newblock {\em npj Computational Materials}, 6(1):20, 2020.

\bibitem{vandermause2022active}
Jonathan Vandermause, Yu~Xie, Jin~Soo Lim, Cameron~J Owen, and Boris Kozinsky.
\newblock Active learning of reactive bayesian force fields applied to
  heterogeneous catalysis dynamics of h/pt.
\newblock {\em Nature Communications}, 13(1):5183, 2022.

\bibitem{xie2023uncertainty}
Yu~Xie, Jonathan Vandermause, Senja Ramakers, Nakib~H Protik, Anders Johansson,
  and Boris Kozinsky.
\newblock Uncertainty-aware molecular dynamics from bayesian active learning
  for phase transformations and thermal transport in sic.
\newblock {\em npj Computational Materials}, 9(1):36, 2023.

\bibitem{hu2022impact}
Taiping Hu, Jianxin Tian, Fuzhi Dai, Xiaoxu Wang, Rui Wen, and Shenzhen Xu.
\newblock Impact of the local environment on li ion transport in inorganic
  components of solid electrolyte interphases.
\newblock {\em Journal of the American Chemical Society}, 145(2):1327--1333,
  2022.

\bibitem{wang2022resistive}
Juefan Wang, Abhishek~A Panchal, Gopalakrishnan~Sai Gautam, and Pieremanuele
  Canepa.
\newblock The resistive nature of decomposing interfaces of solid electrolytes
  with alkali metal electrodes.
\newblock {\em Journal of Materials Chemistry A}, 10(37):19732--19742, 2022.

\bibitem{bekaert2023assessing}
Lieven Bekaert, Suzuno Akatsuka, Naoto Tanibata, Frank De~Proft, Annick Hubin,
  Mesfin~Haile Mamme, and Masanobu Nakayama.
\newblock Assessing the reactivity of the na3ps4 solid-state electrolyte with
  the sodium metal negative electrode using total trajectory analysis with
  neural-network potential molecular dynamics.
\newblock {\em The Journal of Physical Chemistry C}, 127(18):8503--8514, 2023.

\bibitem{ren2024visualizing}
Fucheng Ren, Yuqi Wu, Wenhua Zuo, Wengao Zhao, Siyuan Pan, Hongxin Lin,
  Haichuan Yu, Jing Lin, Min Lin, Xiayin Yao, et~al.
\newblock Visualizing the sei formation between lithium metal and solid-state
  electrolyte.
\newblock {\em Energy \& Environmental Science}, 17(8):2743--2752, 2024.

\bibitem{chaney2024two}
Gracie Chaney, Andrey Golov, Ambroise van Roekeghem, Javier Carrasco, and
  Natalio Mingo.
\newblock Two-step growth mechanism of the solid electrolyte interphase in
  argyrodyte/li-metal contacts.
\newblock {\em ACS Applied Materials \& Interfaces}, 16(19):24624--24630, 2024.

\bibitem{goodwin2024transferability}
Zachary~AH Goodwin, Malia~B Wenny, Julia~H Yang, Andrea Cepellotti, Jingxuan
  Ding, Kyle Bystrom, Blake~R Duschatko, Anders Johansson, Lixin Sun, Simon
  Batzner, et~al.
\newblock Transferability and accuracy of ionic liquid simulations with
  equivariant machine learning interatomic potentials.
\newblock {\em The Journal of Physical Chemistry Letters}, 15(30):7539--7547,
  2024.

\bibitem{larsen2016robust}
Peter~Mahler Larsen, S{\o}ren Schmidt, and Jakob Schi{\o}tz.
\newblock Robust structural identification via polyhedral template matching.
\newblock {\em Modelling and Simulation in Materials Science and Engineering},
  24(5):055007, 2016.

\bibitem{lee2021stack}
Chanhee Lee, Sang~Yun Han, John~A Lewis, Pralav~P Shetty, David Yeh, Yuhgene
  Liu, Emily Klein, Hyun-Wook Lee, and Matthew~T McDowell.
\newblock Stack pressure measurements to probe the evolution of the
  lithium--solid-state electrolyte interface.
\newblock {\em ACS Energy Letters}, 6(9):3261--3269, 2021.

\bibitem{wang2019characterizing}
Michael~J Wang, Rishav Choudhury, and Jeff Sakamoto.
\newblock Characterizing the li-solid-electrolyte interface dynamics as a
  function of stack pressure and current density.
\newblock {\em Joule}, 3(9):2165--2178, 2019.

\bibitem{yang2023lithium}
Menghao Yang, Yunsheng Liu, and Yifei Mo.
\newblock Lithium crystallization at solid interfaces.
\newblock {\em Nature Communications}, 14(1):2986, 2023.

\bibitem{olinger1983lithium}
Bart Olinger and JW~Shaner.
\newblock Lithium, compression and high-pressure structure.
\newblock {\em Science}, 219(4588):1071--1072, 1983.

\bibitem{aktekin2023sei}
Burak Aktekin, Luise~M Riegger, Svenja-K Otto, Till Fuchs, Anja Henss, and
  J{\"u}rgen Janek.
\newblock Sei growth on lithium metal anodes in solid-state batteries
  quantified with coulometric titration time analysis.
\newblock {\em Nature Communications}, 14(1):6946, 2023.

\bibitem{drautz2019atomic}
Ralf Drautz.
\newblock Atomic cluster expansion for accurate and transferable interatomic
  potentials.
\newblock {\em Physical Review B}, 99(1):014104, 2019.

\bibitem{rodriguez2018computing}
Alex Rodriguez, Maria d’Errico, Elena Facco, and Alessandro Laio.
\newblock Computing the free energy without collective variables.
\newblock {\em Journal of chemical theory and computation}, 14(3):1206--1215,
  2018.

\bibitem{dErrico2021automatic}
Maria d’Errico, Elena Facco, Alessandro Laio, and Alex Rodriguez.
\newblock Automatic topography of high-dimensional data sets by non-parametric
  density peak clustering.
\newblock {\em Information Sciences}, 560:476--492, 2021.

\bibitem{glielmo2022dadapy}
Aldo Glielmo, Iuri Macocco, Diego Doimo, Matteo Carli, Claudio Zeni, Romina
  Wild, Maria d’Errico, Alex Rodriguez, and Alessandro Laio.
\newblock Dadapy: Distance-based analysis of data-manifolds in python.
\newblock {\em Patterns}, 3(10), 2022.

\bibitem{Sormani2019}
Giulia Sormani, Alex Rodriguez, and Alessandro Laio.
\newblock {Explicit Characterization of the Free-Energy Landscape of a Protein
  in the Space of All Its C{$\alpha$} Carbons}.
\newblock {\em Journal of Chemical Theory and Computation}, 16(1):80--87, 2020.

\bibitem{Carli2020CandidateSimulations}
Matteo Carli, Giulia Sormani, Alex Rodriguez, and Alessandro Laio.
\newblock {Candidate Binding Sites for Allosteric Inhibition of the SARS-CoV-2
  Main Protease from the Analysis of Large-Scale Molecular Dynamics
  Simulations}.
\newblock {\em The Journal of Physical Chemistry Letters}, 12:65--72, 2020.

\bibitem{Carli2021Interpolation}
Matteo Carli and Alessandro Laio.
\newblock Statistically unbiased free energy estimates from biased simulations.
\newblock {\em Molecular Physics}, 119(19-20):e1899323, 2021.

\bibitem{conder2017direct}
Joanna Conder, Renaud Bouchet, Sigita Trabesinger, Cyril Marino, Lorenz Gubler,
  and Claire Villevieille.
\newblock Direct observation of lithium polysulfides in lithium--sulfur
  batteries using operando x-ray diffraction.
\newblock {\em Nature Energy}, 2(6):1--7, 2017.

\bibitem{yang2021rich}
Jin-Lin Yang, Da-Qian Cai, Xiao-Ge Hao, Ling Huang, Qiaowei Lin, Xiang-Tian
  Zeng, Shi-Xi Zhao, and Wei Lv.
\newblock Rich heterointerfaces enabling rapid polysulfides conversion and
  regulated li2s deposition for high-performance lithium--sulfur batteries.
\newblock {\em ACS nano}, 15(7):11491--11500, 2021.

\bibitem{qi2023electrochemical}
Xiaoqun Qi, Fengyi Yang, Pengfei Sang, Zhenglu Zhu, Xiaoyu Jin, Yujun Pan, Jie
  Ji, Ruining Jiang, Haoran Du, Yongsheng Ji, et~al.
\newblock Electrochemical reactivation of dead li2s for li- s batteries in
  non-solvating electrolytes.
\newblock {\em Angewandte Chemie International Edition}, 62(9):e202218803,
  2023.

\bibitem{sun2016thermodynamic}
Wenhao Sun, Stephen~T Dacek, Shyue~Ping Ong, Geoffroy Hautier, Anubhav Jain,
  William~D Richards, Anthony~C Gamst, Kristin~A Persson, and Gerbrand Ceder.
\newblock The thermodynamic scale of inorganic crystalline metastability.
\newblock {\em Science advances}, 2(11):e1600225, 2016.

\bibitem{hegde2020phase}
Vinay~I Hegde, Muratahan Aykol, Scott Kirklin, and Chris Wolverton.
\newblock The phase stability network of all inorganic materials.
\newblock {\em Science advances}, 6(9):eaay5606, 2020.

\bibitem{chromik1999thermodynamic}
RR~Chromik, WK~Neils, and EJ~Cotts.
\newblock Thermodynamic and kinetic study of solid state reactions in the
  cu--si system.
\newblock {\em Journal of Applied Physics}, 86(8):4273--4281, 1999.

\bibitem{perez2006combined}
LA~P{\'e}rez-Maqueda, JM~Criado, and PE~Sanchez-Jimenez.
\newblock Combined kinetic analysis of solid-state reactions: a powerful tool
  for the simultaneous determination of kinetic parameters and the kinetic
  model without previous assumptions on the reaction mechanism.
\newblock {\em The Journal of Physical Chemistry A}, 110(45):12456--12462,
  2006.

\bibitem{gusak2019phase}
Andriy Gusak, Tetiana Zaporozhets, and Nadiia Storozhuk.
\newblock Phase competition in solid-state reactive diffusion
  revisited—stochastic kinetic mean-field approach.
\newblock {\em The Journal of chemical physics}, 150(17), 2019.

\bibitem{chen2022classical}
Long-Qing Chen and Yuhong Zhao.
\newblock From classical thermodynamics to phase-field method.
\newblock {\em Progress in Materials Science}, 124:100868, 2022.

\bibitem{ziegler1985stopping}
James~F Ziegler and Jochen~P Biersack.
\newblock The stopping and range of ions in matter.
\newblock In {\em Treatise on heavy-ion science: volume 6: astrophysics,
  chemistry, and condensed matter}, pages 93--129. Springer, 1985.

\bibitem{Kresse1993}
G.~Kresse and J.J. Hafner.
\newblock Ab initio molecular dynamics for liquid metals.
\newblock {\em Physical Review B}, 47:558(R), 1993.

\bibitem{KresseFurthPRB}
G~Kresse and J~Furthm{\"u}ller.
\newblock {Efficient iterative schemes for ab initio total-energy calculations
  using a plane-wave basis set}.
\newblock {\em Physical Review B}, 54(16):11169--11186, October 1996.

\bibitem{KresseFurthCMS}
G~Kresse and J~Furthm{\"u}ller.
\newblock {Efficiency of ab initio total energy calculations for metals and
  semiconductors using a plane-wave basis set}.
\newblock {\em Computational Materials Science}, 6(1):15--50, July 1996.

\bibitem{refLDA}
J.P. Perdew and A.~Zunger.
\newblock Self-interaction correction to density-functional approximations for
  many-electron systems.
\newblock {\em Physical Review B}, 23:5048--5079, 1981.

\bibitem{refPBE}
J.P. Perdew, K.~Burke, and M.~Ernzerhof.
\newblock Generalized gradient approximation made simple.
\newblock {\em Physical Review Letters}, 77:3865--3868, 1996.

\bibitem{zuo2020performance}
Yunxing Zuo, Chi Chen, Xiangguo Li, Zhi Deng, Yiming Chen, J{\"o}rg Behler,
  G{\'a}bor Cs{\'a}nyi, Alexander~V Shapeev, Aidan~P Thompson, Mitchell~A Wood,
  et~al.
\newblock Performance and cost assessment of machine learning interatomic
  potentials.
\newblock {\em The Journal of Physical Chemistry A}, 124(4):731--745, 2020.

\bibitem{phuthi2024accurate}
Mgcini~Keith Phuthi, Archie~Mingze Yao, Simon Batzner, Albert Musaelian, Pinwen
  Guan, Boris Kozinsky, Ekin~Dogus Cubuk, and Venkatasubramanian Viswanathan.
\newblock Accurate surface and finite-temperature bulk properties of lithium
  metal at large scales using machine learning interaction potentials.
\newblock {\em ACS omega}, 9(9):10904--10912, 2024.

\bibitem{LAMMPS}
A.~P. Thompson, H.~M. Aktulga, R.~Berger, D.~S. Bolintineanu, W.~M. Brown,
  P.~S. Crozier, P.~J. in~'t Veld, A.~Kohlmeyer, S.~G. Moore, T.~D. Nguyen,
  R.~Shan, M.~J. Stevens, J.~Tranchida, C.~Trott, and S.~J. Plimpton.
\newblock {LAMMPS} - a flexible simulation tool for particle-based materials
  modeling at the atomic, meso, and continuum scales.
\newblock {\em Comp. Phys. Comm.}, 271:108171, 2022.

\end{thebibliography}
\bibliographystyle{unsrt}

\clearpage

\end{document}